\documentclass{sig-alternate}
\usepackage{url}
\usepackage[plainpages=false,pdfpagelabels,
colorlinks=true,citecolor=blue,hypertexnames=false]{hyperref}
\newcommand{\Tx}{\ensuremath{\theta_X}} 
\newcommand{\Dx}{\ensuremath{\partial_X}}
\newcommand{\Dy}{\ensuremath{\partial_Y}} 
\newcommand{\Wr}{\ensuremath{\operatorname{Wr}}}

\newcommand{\M}{\ensuremath{\mathsf{M}}} 
 
\newcommand{\bigO}{{\mathcal{O}}} 
\newcommand{\bigOsoft}{\ensuremath{\tilde{\mathcal{O}}}}

\newcommand{\A}{{\mathbb{A}}} 
\newcommand{\E}{{\mathbb{E}}} 
\newcommand{\Z}{{\mathbb{Z}}} 
\newcommand{\K}{{\mathbb{K}}} 
\newcommand{\KK}{{\overline{\mathbb{K}}}} 
\newcommand{\LL}{{\mathbb{L}}} 
\newcommand{\Q}{{\mathbb{Q}}}

\newcommand{\C}{{\mathbb{C}}}

\newtheorem{lemma}{Lemma}
\newtheorem{theorem}{Theorem}

\newtheorem{proposition}{Proposition}
\newtheorem{definition}{Definition}
\newtheorem{hypothesis}{Hypothesis} 

\begin{document} 
\conferenceinfo{ISSAC'07} {July 29-August 1, 2007, Waterloo, Ontario,
  Canada.}
\CopyrightYear{2007}
\crdata{978-1-59593-743-8/07/0007}
\title{Differential Equations for Algebraic Functions}
\newfont{\eaddfntsmall}{phvr at 10pt}
%
%
\def\more-auths{%
\end{tabular}
\begin{tabular}{c}}
\numberofauthors{2}
\author{
  \alignauthor 
  \hbox{Alin Bostan, Frédéric Chyzak, Bruno Salvy}
  \affaddr{Algorithms Project, Inria Rocquencourt}
  \affaddr{78153 Le Chesnay (France)}
  \email{\eaddfntsmall {\{Alin.Bostan,Frederic.Chyzak,Bruno.Salvy\}@inria.fr}}
  \more-auths
  \alignauthor Grégoire Lecerf\\
  \affaddr{Laboratoire de Math\'ematiques (CNRS)} 
  \affaddr{Universit\'e de Versailles St-Quentin-en-Yvelines} 
  \affaddr{78035 Versailles (France)}
  \email{\eaddfntsmall {lecerf@math.uvsq.fr}}
  \alignauthor {\'E}ric Schost\\
  \affaddr{Department of Computer Science}
  \affaddr{University of Western Ontario}
  \affaddr{London, Ontario (Canada)}
  \email{\eaddfntsmall {eschost@uwo.ca}}
}
\maketitle           
\begin{abstract}   
  It is classical that univariate algebraic functions satisfy linear
  differential equations with polynomial coefficients. 
  Linear recurrences follow for the coefficients of their power series
  expansions. We show that the linear differential equation of minimal
  order has coefficients whose degree is cubic in the degree of the
  function. We also show that there exists a linear differential
  equation of order linear in the degree whose coefficients are only
  of quadratic degree. Furthermore, we prove the existence of
  recurrences of order and degree close to optimal. We study the
  complexity of computing these differential equations and
  recurrences. We deduce a fast algorithm for the expansion of algebraic series.
\end{abstract} 
\par
\bigskip
\noindent
{\bf Categories and Subject Descriptors:} I.1.2 {[{\bf Symbolic and
    Algebraic Manipulation}]}: {Algorithms}
\par
\vspace{1mm}
\noindent
{\bf General Terms:} Algorithms, Experimentation, Theory
\par
\vspace{1mm}
\section{Introduction}
A power series~$\alpha(X)\in\K[[X]]$ is called algebraic over the
field~$\K$ when there is a nonzero polynomial~$P\in\K[X,Y]$ such that
$P(X,\alpha(X))=0$. 

The existence of a linear differential equation with polynomial
coefficients satisfied by algebraic series seems to have been observed
first in~1827 by Abel~\cite[p.~287]{Abel92} who did not publish
it. Cockle gave an algorithm for the computation of such a
differential equation, that he called a \emph{differential
  resolvent}~\cite{Cockle1861}. The method was then presented by
Harley~\cite{Harley1862} and rediscovered by
Tannery~\cite{Tannery1875} (see also~\cite{Gray86}). 

One of the applications of these differential equations is the
computation of power series solutions: a linear differential equation
translates into a linear recurrence, with the consequence that the
number of operations required to compute the first~$N$ coefficients
grows only linearly with~$N$. This is useful in combinatorial
applications, since the enumeration sequences of context-free
languages have algebraic generating series. This method has been
described by Comtet \cite{Comtet64} and studied from the complexity
point of view by Chudnovsky and Chudnovsky~\cite{ChCh86}. In
practice, this procedure will be all the more efficient as its
dependency on the size of~$P$ is small. This in turn is related to the
order and degree of the coefficients of the differential equation. 

We say that a linear differential equation or operator vanishing at
all the roots~$\alpha(X)$ of~$P$ is \emph{associated to~$P$}. In this
work, we derive precise bounds for the order and degree of
coefficients of several such equations. For brevity and clarity, the
following hypothesis will be assumed throughout.
\begin{hypothesis}[H]
  The characteristic of the field~$\K$ is~0. The polynomial $P\in\K[X,Y]$ is
  \emph{separable} (wrt~$Y$), i.e., coprime with its
  derivative~$P_Y$ wrt~$Y$. The total degree of~$P$ is denoted~$D$
  and its degrees in~$X$ and~$Y$ are written $D_X$ and $D_Y$.
  We assume that~$D_Y\ge 1$.
\end{hypothesis}
The first work on bounds that we know of in this area is due to
Cormier \emph{et alii}~\cite{CoSiTrUl02}. They consider the monic
linear differential operator of minimal order associated to~$P$, which 
we call \emph{the differential resolvent of~$P$}. A
bound in~$\bigO(D_X D_Y^4)$ for the degrees of its coefficients can be
deduced from their work.
Nahay~\cite{Nahay03, Nahay04} 
obtained degree bounds in~$\bigO(D^3)$ for the coefficients of an analogous
equation satisfied by~$\alpha^\lambda$ with~$\lambda$ transcendent
over~$\K$. 
In \S\ref{resolvent}, we prove the following cubic bound ($\Dx$
denotes differentiation wrt~$X$).
\begin{theorem}\label{tm:cockle}
  Under (H), the differential resolvent of~$P$ can be written
  $$
  \Dx^r + \frac{M_{r-1}}{M_r} \Dx^{r-1} + \dots + \frac{M_1}{M_r} \Dx
  + \frac{M_0}{M_r},
  $$
  where $r\le D_Y$ and $M_0,\ldots,M_r$ in $\K[X]$ have degrees at
  most
  $$
  \eta= ((2r-1)D_Y+2r^2-4r+3)D_X-r(r-1)/2 \in \bigO( r D_X D_Y).  $$
\end{theorem}
Experiments show that a cubic degree is to be expected (see
\S\ref{optimal-bounds}). This theorem improves and extends the
results of~\cite{CoSiTrUl02,Nahay03,Nahay04}. It leads to a linear
recurrence of order~$\bigO(D^3)$ with coefficients of
degree~$\bigO(D)$.  Using fast multipoint evaluation on an arithmetic
progression then leads to a computation of the first~$N$ coefficients
in~$\bigO(D^2\M(D)N)$ arithmetic operations (see \S\ref{recurrence}).
As usual, $\M(D)$ denotes a bound on the number of arithmetic
operations in the ring of coefficients needed to multiply two
polynomials of degree at most~$D$. We make the standard assumption
that~$\M(d_1)+\M(d_2)\le \M(d_1+d_2)$ for all 
$d_1,d_2$ and that~$D\log D \in \bigO(\M(D))$.
\par
One reason for this cubic degree growth is the presence of numerous
\emph{apparent singularities}. These are points where the differential
resolvent is singular (its leading coefficient vanishes) while none of
its solutions is.  However, the number of apparent singularities
decreases dramatically if one does not insist on
the order of the equation being minimal. It is actually possible to
remove all apparent singularities by \emph{Weyl
  closure}~\cite{Tsai00b}, but we do not know how to get bounds from
this algorithm.
\par
In~\S\ref{simple-bounds}, we show that a moderate increase of the
order leads to the following \emph{quadratic} bound.
\begin{theorem}\label{th:simple-bound}
  Under (H), there exists a nonzero linear differential operator
  in~$\K[X]\langle\Dx\rangle$ associated to~$P$ whose degree in~$\Dx$
  is at most~$6D_Y$ and with coefficients of degree at most $3D_X
  D_Y$.
\end{theorem}
This operator then translates into a linear recurrence of order at
most~$3D_Y(D_X+2)$ with coefficients of degree at most~$6D_Y$. Then, a
natural question is the order of a minimal linear recurrence, or
equivalently the minimal degree in~$X$ of linear differential
operators in~$X$ and~$\theta_X=X\partial_X$ associated to~$P$.
In~\S\ref{optimal-bounds}, we explain why quadratic order is optimal.
We prove in \S\ref{creative-telescoping} the following precise bound.
\begin{theorem}\label{thm:bidegree-bound}
  Under (H), there exists a non-zero differential operator in
  $\K[X]\langle\theta_X\rangle$ associated to~$P$ whose degrees in~$X$
  and in~$\theta_X$ are bounded above by
  \[2D_XD_Y+D_Y-\Delta^2-\Delta
  +1\quad\text{with}\quad\Delta=D_X+D_Y-D.\]
\end{theorem}
Experiments showing that this bound is not far from optimal are
reported in~\S\ref{optimal-bounds}.  This result implies the existence
of a linear recurrence of order~$\bigO(D^2)$ with coefficients of the
same degree, whence a computation of the first~$N $ coefficients of
series solutions in~$\bigO(\M(D^2)N)$, with a small constant in
the~$\bigO(\cdot)$ term.
%
%
%
In~\S\ref{efficient-algo}, we go a bit further: with a degree in~$X$ slightly larger, but still quadratic, we obtain a degree in~$\theta_X$ that is only linear; this results in an algorithm for algebraic series in~$\bigO(D\M(D)N)$ operations only.

%
\section{Differential Resolvent} \label{resolvent}
We write $\alpha_1,\ldots,\alpha_{D_Y}$ for the zeroes of~$P(X,\cdot)$
in the algebraic closure~$\overline{\K(X)}$ of~$\K(X)$. Since $P$ is
{separable} wrt~$Y$,
the derivation $\Dx$ uniquely extends to the splitting field
$\K(X)(\alpha_1,\ldots,\alpha_{D_Y})$ of $P$ by
$\Dx\alpha_i=-P_X(X,\alpha_i)/P_Y(X,\alpha_i)$.
%
%
\subsection{Cockle's Algorithm}   
Cockle's algorithm~\cite{Cockle1861} computes the differential
resolvent.  It reduces the computation to linear algebra in the
quotient ring $\K(X)[Y]/(P)$, that is a vector space of
dimension~$D_Y$ over $\K(X)$.  The heart of the idea is to express the successive derivatives of any~$\alpha_i$ as a polynomial of degree at most~$D_Y-1$ in itself. In order to prepare the derivation of bounds, we
introduce two sequences of polynomials.

First, polynomials $(W_k)_{k\ge1}$ in $\K[X,Y]$ 
satisfying
\[\alpha_i^{(k)} =
\frac{W_k(X,\alpha_i)}{P_Y(X,\alpha_i)^{2k-1}},\quad i \in
\{1,\ldots,D_Y\}\] are defined by induction: 
$W_1 =
-P_X$, and for all $k\ge 1$, $W_{k+1} = ( P_Y \Dx W_k - P_X \Dy W_k )
P_Y - (2k-1) ( P_Y \Dx P_Y - P_X \Dy P_Y) W_k$. They satisfy
\begin{align*}                 
  \deg_X(W_k) &\le (2 D_X -1)k - D_X,\\
  \deg_Y(W_k) &\le 2(D_Y-1)k -D_Y+2.
\end{align*}
Cockle's algorithm computes a second sequence of polynomials,
$(V_k)_{k\ge 0}$, such that~$V_k(\alpha)$ is the image of $W_k / P_Y^{2k-1}$
in $\K(X)[Y]/(P)$ for $k\ge1$, while~$V_0(\alpha)=\alpha$ is that of~$Y$. The
computation is as follows:
\begin{enumerate} 
\item $V_1$ is computed by first inverting $P_Y$ modulo $P$ and then
  reducing the product with $-P_X$ modulo
  $P$;
\item for $k=0,1,\dotsc$, the polynomial~$V_{k+1}$ is computed by taking the remainder of 
  $\Dx V_k+V_1\Dy V_k$ modulo $P$;
\item the computation stops with the first integer $r\le D_Y$ such
  that $V_r$ belongs to the $\K(X)$-vector space spanned by
  $V_0,\ldots,V_{r-1}$. Then, together with $r$, Cockle's algorithm
  computes the relation
  \begin{equation}\label{linear-relation}
    V_r = A_{r-1} V_{r-1} + \dots + A_0 V_0,
  \end{equation}
  and returns the unique monic differential operator of minimal order
  $M = \Dx^r - A_{r-1} \Dx^{r-1} - \dots - A_1 \Dx - A_0$ associated
  to $P$.
\end{enumerate}
%
%
\subsection{Degree Bound}   
%
%
%
We are now going to prove Thm.~\ref{tm:cockle} by establishing a
formula for the differential resolvent in terms of the elementary
symmetric polynomials of the $\alpha_i$. This idea originates in
Nahay's PhD thesis~\cite[Thm.~46, p.~96]{Nahay00} (published
in~\cite[Thm.~9.1]{Nahay04}).
\par
Let $Y_1,\ldots,Y_{D_Y}$ be new indeterminates and  ${\mathcal N}$  the
matrix:
$${\scriptsize
  \left(
    \begin{matrix}
      Y_1 P_Y(X,Y_1)^{2r-1} & \cdots & Y_{r}
      P_Y(X,Y_{r})^{2r-1} & 1 \\
      W_1(X,Y_1) P_Y(X,Y_1)^{2r-2} & \cdots & W_1(X,Y_{r})
      P_Y(X,Y_{r})^{2r-2} & \Dx \\
      W_2(X,Y_1) P_Y(X,Y_1)^{2r-4} & \cdots & W_2(X,Y_{r})
      P_Y(X,Y_{r})^{2r-4} & \Dx^2 \\
      \vdots & & \vdots & \vdots\\
      W_{r}(X,Y_1) & \cdots & W_{r}(X,Y_{r})
      & \Dx^r \\
    \end{matrix}
  \right).}
$$
By subtracting the $j$th column of ${\mathcal N}$ to the $i$th one
($i<j \le r$) we observe that $Y_i - Y_j$ divides the determinant
$\det({\mathcal N})$ of ${\mathcal N}$, hence $\delta :=
\prod_{1 \le i < j \le r}(Y_i - Y_j)$ divides
$\det({\mathcal N})$. The operator $L = \det({\mathcal N})/\delta$ thus belongs to
$\K[X][Y_1,\ldots,Y_r]\langle\Dx\rangle$ and it is symmetric in
$Y_1,\ldots,Y_r$.
\par
By definition of the $W_k$'s, for all $\sigma$ in the permutation
group $S_{D_Y}$ of $\{1,\ldots,D_Y\}$ and all $i\in\{1,\dots,r\}$, we
have
\begin{multline*}  L(\alpha_{\sigma(1)},\dots,\alpha_{\sigma(r)},\Dx)
  \alpha_{i}=\\ \left(\prod_{j=1}^r P_Y(X,\alpha_{\sigma(j)})
  \right)^{2r-1} \frac{\Wr(\alpha_{\sigma(1)},\ldots,
    \alpha_{\sigma(r)},\alpha_i)}{\delta(\alpha_{\sigma(1)},\ldots,
    \alpha_{\sigma(r)})} =0,
\end{multline*}
where $\textrm{Wr}(\cdot)$ is the Wronskian determinant. In other
words $L(\alpha_{\sigma(1)},\ldots,\alpha_{\sigma(r)},\Dx)$ is
associated to $P$ for all $\sigma \in S_{D_Y}$.
\par
For all $k \in \{0,\ldots,r\}$, the entries of the $(k+1)$th row of
${\mathcal N}$ have degree in $X$ at most $m_k = (2r-1)D_X -k$,
except in the last column. Therefore, for all $i \in \{0,\ldots,r\}$,
the degree of the coefficient $L_i$ of $\Dx^{i}$ in $L$ is bounded by:
%
\[    \deg_X(L_i) \le  \sum_{k = 0,k \ne i}^r m_k 
\le r(2r-1)D_X -r(r-1)/2  =: m_X.
\]
%
On the other hand, for all $l \in \{1,\ldots,r\}$, the entries of the
$l$th column of ${\mathcal N}$ have degree in $Y_l$ at most $(2r-1)D_Y
-2r +2$. Therefore, for all $i \in \{0,\ldots,r\}$, all $j \ge 0$, and
all $l \in \{1,\ldots,r\}$, the degree of the coefficient $L_{i,j}$ of
$X^j$ in $L_i$ is bounded by:
%
\[ \deg_{Y_l}(L_{i,j}) \le (2r-1)D_Y - 2r +2 - (r-1)
=: m_Y.
\]
By definition of $r$, there exists a permutation $\sigma \in S_{D_Y}$
such that $L(\alpha_{\sigma(1)}, \ldots, \alpha_{\sigma(r)},\Dx)$ has
order $r$ exactly. Therefore, Lemma~\ref{lm:cocklea} below provides us
with a nonzero operator $\bar{L}
\in\K[X][Y_1,\ldots,Y_{D_Y}]\langle\Dx\rangle$ symmetric in
$Y_1,\ldots,Y_{D_Y}$, such that $\bar{L}(\alpha_1,\ldots,
\alpha_{D_Y},\Dx)$ has order $r$, is associated to $P$, and 
$\deg_{Y_i}(\bar{L}) \le \max(m_Y, D_Y - 1) = m_Y$, for all $i\in
\{1,\ldots,D_Y\}$.
\par
Let $p_i$ denote the coefficient of $Y^i$ in $P$ seen in $\K[X][Y]$.
By Lemma~\ref{lm:cockleb} below, there exists $\tilde{L} \in
\K[X][Z_1,\ldots,Z_{D_Y}]\langle\Dx\rangle$ of total degree at most
$m_Y$ in $Z_1,\ldots,Z_{D_Y}$, of partial degree in $X$ at most $m_X$,
and such that $\bar{L}(\alpha_1,\ldots,\alpha_{D_Y},\Dx) =
\tilde{L}(p_0/p_{D_Y},\ldots, p_{D_Y-1}/p_{D_Y},\Dx)$.
\par
Finally, $\tilde{M} = p_{D_Y}^{m_Y} \tilde{L}(p_0/p_{D_Y},\ldots,
p_{D_Y-1}/p_{D_Y},\Dx)$ is in $\K[X]\langle\Dx\rangle$,  
has order $r$ and degree at most $m_X + D_X m_Y$ in $X$, 
and is
associated to $P$. The conclusion of Thm.~\ref{tm:cockle} thus follows from the uniqueness of
the minimal operator.
\par
\medskip The rest of the proof consists of elementary properties of
symmetric polynomials.
If $r \le n$ then the group $S_r$ of  permutations of
$\{1,\ldots,r\}$ is naturally embedded into $S_n$. We write
$(S_n/S_r)_l$ for the left quotient of $S_n$ by $S_r$.
\begin{lemma}\label{lm:cocklea}
  Let $\E$ be a field, let $r \le n$ be nonnegative integers, and
  let $\alpha_1,\ldots,\alpha_n$ be pairwise distinct elements in
  $\E$. If $E \in \E[Y_1,\ldots,Y_r]$ is symmetric and does not vanish
  at $(\alpha_1,\ldots,\alpha_r)$, then there exist nonnegative
  integers $e_{r+1} \le r,\ldots,e_n \le n-1$ such that
  $$
  \bar{E} = \sum_{\sigma \in (S_n/S_r)_l}
  E(Y_{\sigma(1)},\ldots,Y_{\sigma(r)}) Y_{\sigma(r+1)}^{e_{r+1}}
  \cdots Y_{\sigma(n)}^{e_{n}}
  $$
  is a symmetric polynomial in $\E[Y_1,\ldots,Y_n]$ that does not
  vanish at $(\alpha_1,\ldots,\alpha_n)$.
\end{lemma}
\begin{proof}
  For fixed $r$, the proof proceeds by induction on~$n$. If
  $n=r$, then the lemma trivially holds. We now assume that the
  lemma holds for some $n\ge r$, and prove that it holds for
  $n+1$. The induction hypothesis provides us with nonnegative
  integers $e_{r+1} \le r,\ldots,e_n \le n-1$ such that $\bar{E}$ (as
  defined in the lemma) is a symmetric polynomial in $Y_1,\ldots,Y_n$
  that does not vanish at $(\alpha_1,\ldots,\alpha_n)$.  Since the
  Vandermonde matrix of $\alpha_1,\ldots,\alpha_{n+1}$ is invertible,
  there exists $e_{n+1} \le n$ such that
  $$
  \sum_{\rho \in (S_{n+1}/S_n)_l}
  \bar{E}(Y_{\rho(1)},\ldots,Y_{\rho(n)}) Y_{\rho(n+1)}^{e_{n+1}}
  $$
  is symmetric in $Y_1,\ldots,Y_{n+1}$ and moreover does not vanish at
  $(\alpha_1,\ldots,\alpha_{n+1})$. Plugging the expression of $\bar{E}$
  into the latter sum concludes the proof for $n+1$.
\end{proof}
\begin{lemma}\cite[Thm~6.21]{Yap00}\label{lm:cockleb}
  Let $\A$ be a commutative ring, and let $E \in \A[Y_1,\ldots,Y_n]$ be
  symmetric of partial degree $D$ in each of the $Y_i$'s. Then there
  exists a unique polynomial $F \in \A[Z_1,\ldots,Z_n]$ of total degree
  $D$ such that $E(Y_1,\ldots,Y_n) = F(\Sigma_1,\ldots,\Sigma_n)$,
  where $\Sigma_i$ represents the $i$-th elementary symmetric
  polynomial in $Y_1,\ldots,Y_n$.
\end{lemma}
\subsection{Algorithm and Complexity}\label{algo-cockle}
With the bounds of the previous section, we now turn to the
computation of the differential resolvent. As
in~\cite{ChCh86,CoSiTrUl02}, the idea is to compute power series
expansions of the coefficients and recover them by Pad\'e
approximants. For this, we have to choose a point where expansion is
easy and determine a bound on the order of the series expansions.
\begin{definition}
  A point~$a\in\K$ that annihilates neither the leading coefficient
  $p_{D_Y}$ nor the discriminant $\Delta_Y$ of $P$ seen in $\K[X][Y]$,
  and such that $V_0(a),\ldots,V_{r-1}(a)$ are linearly independent in
  $\K[Y]/(P(a,Y))$ is called a \emph{lucky} point.
\end{definition}
If $a$ is lucky, then the $\alpha_i$'s can be seen as power series in
$\KK[[X-a]]$, and the $V_i$'s as elements of $\K[[X-a]][Y]/(P(a,Y))$. Here
$\KK$ represents the algebraic closure of $\K$.
\begin{lemma}
  \label{lemma4}
  There exists $\Phi \in\K[X] \setminus \{0\}$ of total degree at most
  $\eta$ (as defined in Thm.~\ref{tm:cockle}) such that $a$ is
  lucky whenever $p_{D_Y}(a) \Delta_Y(a) \Phi(a) \ne 0$.
\end{lemma}
\begin{proof}
  For $\Phi$ we take the leading coefficient of the operator
  $\tilde{M}$ introduced at the end of the proof of
  Thm.~\ref{tm:cockle}. If $p_{D_Y}(a) \Delta_Y(a) \ne 0$ then we can
  define $A(a)$ as the $r \times D_Y$ matrix $(
  \alpha_j^{(i-1)}(a))_{i,j}$ with entries in~$\KK[[X-a]]$, $B(a)$ as the $D_Y \times r$ matrix in
  $\K[[X-a]]$ whose columns are the coefficients in~$Y$ of $V_0,\ldots,V_{r-1}$, and $C(a)$ as the
  invertible Vandermonde matrix of the $\alpha_i$. If $\Phi(a) \ne 0$
  then $A(a)$ has rank $r$ by construction, and so has $B(a)$ since
  $A(a) = C(a) B(a)$.
\end{proof}
Lemma~\ref{lemma4} implies that all but a finite set of elements in
$\K$ are lucky. A lucky point $a$ can thus be obtained either
deterministically by computing the rank of $V_0(a),\ldots,V_{r-1}(a)$
for more than $\deg_X(p_{D_Y} \Delta_Y \Phi) \in \bigO(D_X D_Y^2)$
points, or probabilistically by means of the Schwartz-Zippel zero
test \cite[Lemma~6.44]{GaGe03}.
Once a lucky point is determined (which is very fast by the
probabilistic approach) then the computation of the differential
resolvent of $P$ can be done efficiently.
\begin{proposition}\label{pp:cocklecost}
  Given a lucky point $a \in\K$ and the order~$r$ of $M$, the
  computation of $M$ can be done with $\bigO( (r^\omega + r\M(D_Y) +r
  \log(D_X) )\M(\eta) )$ arithmetic operations in $\K$.
\end{proposition}
Here $2<\omega\le3$ is the exponent of matrix multiplication, see,
e.g., \cite{GaGe03}.
\begin{proof}
  First of all, shifting the variable $X$ by $a$ in $P$ takes
  $\bigO(D_Y \M(D_X))$ operations in $\K$ by means of~\cite[Ch.~1,
  \S2]{BiPa94} since the characteristic is zero. In order to
  reconstruct the coefficients of $M$ from their series expansions via
  Pad\'e approximants~\cite[\S5.9]{GaGe03}, a precision
  $(X-a)^{2\eta+1}$ is sufficient. The idea is to follow the steps of
  Cockle's algorithm in~$\K[[X-a]][Y]/(P)$ instead of~$\K(X)[Y]/(P)$.
  The precision of the series is initially set to $2\eta+1 +r \in
  \bigO(\eta)$ in order to accommodate the precision loss due to
  differentiation.
  \par
  In Step~1, the computation of $V_1$ takes $\bigO(\M(D_Y)\M(\eta))$
  operations by~\cite[Thms.~15.10 \& 15.11]{GaGe03}.  In Step~2, each
  $V_{k+1}$ can be obtained from $V_k$ and $V_1$ by means of
  $\bigO(\M(D_Y)\M(\eta))$ operations.  In Step~3, we start by
  computing the approximation of the relation~\eqref{linear-relation}
  to precision $(X-a)$. This boils down to linear algebra over $\K$ in
  size $(r+1) \times D_Y$, hence a cost in $\bigO(r^{\omega-1} D_Y)$.
  From this we use Hensel-Newton lifting in order to recover
  relation~\eqref{linear-relation} to precision $(X-a)^{2\eta+1}$ with
  $\bigO(r^\omega \M(\eta))$ operations in $\K$. The rational
  reconstruction of each $A_i$ and its shift back of $X$ by $-a$ cost
  $\bigO(\M(\eta)\log(\eta))$.
\end{proof}
When $r = D_Y$ the cost of the computation of $M$ drops to $\bigOsoft(
D_X D_Y^{\omega+2})$ when fast multiplication is used. (Here the tilde
indicates that the factors polynomial in $\log D_X$, $\log D_Y$ and
their logarithms have been omitted.) Note that the results presented
in this section extend to analyze the bit-complexity via $p$-adic
approximation when $\K = \Q$.
\section{Creative Telescoping\\
  for Algebraic Functions}
\label{creative-telescoping}
In this section we search for differential operators associated to~$P$ in $\K(X)\langle
\Dx \rangle$ with non-minimal orders but with
degrees in~$X$ much smaller than in the resolvent of~$P$.
\subsection{Principle}
Let~$\alpha(X)$ be a solution of~$P(X,\alpha(X))=0$. If $P\in\C[X,Y]$,
then by the residue theorem,
\[\alpha(X)=\frac{1}{2\pi i}\oint F(X,Y)dY,\text{ with }
F(X,Y)=\frac{YP_Y(X,Y)}{P(X,Y)}\] where the contour is for instance a
circle containing~$\alpha(X)$ and no other root of~$P$. In this
setting, \emph{creative telescoping} can be seen as an algorithm based on
differentiating under the integral sign and integrating by
parts~\cite{AlZe90}. It consists in computing a linear differential
operator of the form
\begin{equation}\label{eq-lambda}
  \Lambda=A(X,\partial_X)+\partial_YB(X,\partial_X,Y,\partial_Y)
\end{equation}
that cancels the rational function~$F$. There, $\partial_X$ and
$\partial_Y$ denote differentiation with respect to~$X$ and~$Y$. If
such a~$\Lambda$ exists, integrating~$\Lambda(YP_Y/P)=0$ shows that
$A$ cancels~$\alpha$. We now give an algebraic proof of this property.
\begin{proposition}
  \label{prop:creative-telescoping}
  Under (H), suppose that an element 
  $\Lambda\in\K[X,Y]\langle\Dx,\Dy\rangle$ is of the
  form~\eqref{eq-lambda} and that $\Dy^k\Lambda$ cancels the rational
  function~$F=YP_Y/P$ for some integer $k\ge0$.  Then, the
  operator~$A$ is associated to~$P$.
\end{proposition}
%
%
\begin{proof}
  The element~$F=YP_Y/P$ of~$\K(X,Y)$ can be expanded as a Laurent
  series in~$\K(X,\alpha)((Y-\alpha))$:
  \[F=\frac{\alpha}{Y-\alpha}+\sum_{i\ge0}{\gamma_i(Y-\alpha)^i}\]
  with coefficients~$\gamma_i$ in~$\K(X,\alpha)$. On the other hand
  the action of derivations in $X$ and~$Y$ on a summand
  $\gamma(Y-\alpha)^i$, for $\gamma$ in~$\K(X,\alpha)$, is given for
  any~$i\in\Z$ by
  \begin{gather*}
    \partial_Y\left(\gamma(Y-\alpha)^i\right) = i\gamma(Y-\alpha)^{i-1}\\
    \partial_X\left(\gamma(Y-\alpha)^i\right) =
    \gamma'(Y-\alpha)^i-i\gamma(Y-\alpha)^{i-1}\alpha'.
  \end{gather*}
  Thus no term in~$(Y-\alpha)^{-1}$ remains after action
  of~$\partial_Y$. In this way, applying $\Dy^k\Lambda$ to~$F$ and
  considering the resulting Laurent series yields
  \[0=\Dy^k\Lambda(F)=\Dy^k\Lambda\left(\frac\alpha{Y-\alpha}\right)
  \Rightarrow(-1)^k\frac{A(X,\partial_X)(\alpha)}{(Y-\alpha)^{k+1}}=0,\] which proves
  that $A$ annihilates~$\alpha$.
\end{proof}
%
%
\subsection{Quadratic Bound}   
\label{simple-bounds}
We now prove Thm.~\ref{th:simple-bound} by considering the degrees of
the numerators and denominators of the successive derivatives of the
rational
function~$F=Y{P_Y}/{P}$.
%
%
By Leibniz's rule, for any nonnegative integers $N_X$ and
$N_\partial$, the $r_{N_X,N_\partial}:= (N_X+1) \binom{N_\partial +2}{2} $ rational functions
$$ X^i \Dx^j \Dy^k F,  \quad 0 \le i \le N_X,\ 0 \le j+k \leq N_\partial $$
are contained in the $\K$-vector space of dimension at most
$s_{N_X,N_\partial}:= (D_Y(N_\partial+1)+1) (N_X+1 + D_X(N_\partial
+1)) $ spanned by the elements
$$\frac{X^i Y^j}{P^{N_\partial+1}},\quad  0\le i \le N_X + D_X(N_\partial
+1),\ 0\le j \leq D_Y(N_\partial+1).$$ With the values of~$N_X$ and
$N_\partial$ given in the theorem, the difference
$r_{N_X,N_\partial}-s_{N_X,N_\partial}$ becomes
\[(12D_Y^2-7D_Y-1)D_X+12D_Y^2+8D_Y\] which is positive for positive
$D_Y$.
\par
This proves the existence of a nonzero differential operator
$\Lambda(X,\Dx,\Dy)$ with degree in $X$ at most $N_X$ and total degree
in $\Dx$ and $\Dy$ at most $N_\partial$ that annihilates $F$.  Let
now~$k\le N_\partial$ be such that~$\Lambda$ rewrites
$\Dy^k(A(X,\Dx)+\Dy B(X,\Dx,\Dy))$ with~$A\neq0$. The conclusion
follows from Prop.~\ref{prop:creative-telescoping}.
\subsection{Refined Bound} 
\label{sec:degree-bound}
We now prove Thm.~\ref{thm:bidegree-bound}. The bound of the previous
section relies on the elimination of~$Y$ in the left ideal
annihilating~$F$ in the Weyl algebra~$W_{X,Y}(\K)$. In the context of
creative telescoping, this is known as Zeilberger's slow algorithm,
which relies on holonomic D-modules~\cite{Zeilberger91}. 
A faster algorithm for hyperexponential integrands~\cite{AlZe90}
consists in looking for an operator of the form~\eqref{eq-lambda}.
Indeed, letting~$B$ in~\eqref{eq-lambda} depend on~$Y$ also allows for the existence of operators of
smaller order. In this context, the only good bounds~\cite{ApZe06} we know
of do not apply to rational integrands. Instead, we give
a construction inspired by one of Takayama~\cite[Thm.~2.4 and
Appendix~6]{Takayama92} showing that any differentiably finite
function is holonomic.
%
\par
\medskip
For each integer~$d\ge0$, we consider the vector space~${\mathcal V}_d$ of
rational functions~$G/P^{d+1}$, with polynomials~$G\in\K[X,Y]$
satisfying
\begin{gather*}
  \deg G\leq(d+1)D+d,\quad
  \deg_XG\leq(d+1)D_X+d,\\
  \deg_YG\leq(d+1)D_Y.
\end{gather*}
This is defined in such a way that an induction on~$d$ shows that the
vector space~$F_d$ generated by the rational functions
\begin{equation}\label{gen-Fd}
  X^i\theta_X^j\cdot F,\quad 0\le i,j\le d
\end{equation}
is a subspace of~${\mathcal V}_d$, and that so is~$\partial_Y\cdot {\mathcal V}_{d-1}$. We
are going to establish that for~$d$ large enough,
\begin{equation}
  \label{ineq-dim}
  \dim_\K F_d+\dim_\K({\mathcal V}_d\cap\partial_Y\cdot {\mathcal V}_{d+1})>\dim_\K {\mathcal V}_d.
\end{equation}
{}From here the result follows: both~$F_d$ and~${\mathcal V}_d\cap\partial_Y\cdot
{\mathcal V}_{d+1}$ are subspaces of~${\mathcal V}_d$ whose dimensions are such that they
have a nonzero intersection. Any element of this intersection is of
the form~$AF=\partial_Y(G/P^{d+2})$ with~$AF\in F_d$. By the same
arguments as in the proof of Prop.~\ref{prop:creative-telescoping}, we
deduce that $A$ is associated to $P$.
\par
We now consider the dimensions in~\eqref{ineq-dim} more precisely.
\par
The generators~\eqref{gen-Fd} are linearly independent since
$P_Y\neq0$: the reduced form of~$\theta_X^i\cdot F$ has
denominator~$P^{i+1}$, which cannot be eliminated by multiplication by
a polynomial in~$X$ only. Therefore~$\dim_\K F_d=(d+1)^2$.
\par
Write~$P$ in the form~$X^{D_X}\phi+ \tilde{P}$ where $\phi$~depends
only on~$Y$. For~$d\ge0$, define the vector space~${\mathcal W}_d$ of rational
functions~$F/P^d$, with polynomials~$F\in\K[X,Y]$ of the
form~$F=cX^{(D_X+1)d+1}\phi^d+H$ where $c$~is in~$\K$ and
$H$~satisfies
\begin{gather}
  \label{eq:def-Wd-begin}
  \deg H\leq(D+1)d+1,\quad
  \deg_XH\leq(D_X+1)d,\\
  \label{eq:def-Wd-end}
  \deg_YH\leq D_Yd.
\end{gather}
The decomposition of~$F$ induces ${\mathcal W}_d\subseteq {\mathcal V}_{d+1}$ directly.
Then, from
\begin{gather*}
  P^{d+1}\Dy\left(\frac{F}{P^d}\right) = cd \phi^{d-1} X^{(D_X+1)d+1}
  (\tilde{P} \Dy \phi - \phi \Dy\tilde{P})\\
  + P\Dy H - d H \Dy P,
\end{gather*}
we obtain the inclusion~$\partial_Y\cdot {\mathcal W}_d\subseteq {\mathcal V}_d$.  It
follows that $\partial_Y\cdot {\mathcal W}_d\subseteq {\mathcal V}_d\cap\partial_Y\cdot
{\mathcal V}_{d+1}$, and that Eq.~\eqref{ineq-dim}~holds as soon as
\begin{equation}
  \label{ineq-dimb}
  \dim_\K F_d+\dim_\K\partial_Y\cdot {\mathcal W}_d>\dim_\K {\mathcal V}_d.
\end{equation}
In order to compute the dimension of~$\partial_Y\cdot {\mathcal W}_d$, we first
consider the kernel of~$\partial_Y|_{{\mathcal W}_d}$, which is
precisely~$\K[X]\cap {\mathcal W}_d$.  In view of the degree
constraints~(\ref{eq:def-Wd-begin}--\ref{eq:def-Wd-end}), this
corresponds to numerators~$F=P^dQ$ with $Q\in\K[X]$ and
\[\deg_XQ\leq(D_X+1)d+1-D_Xd=d+1,\]
so that $\dim_\K(\partial_Y\cdot {\mathcal W}_d)=\dim_\K {\mathcal W}_d-(d+2)$.
\par
The dimensions of ${\mathcal V}_d$ and~${\mathcal W}_d$ are given by the possible supports
of the numerators of their elements.
Lemma~\ref{lem:combinatoric-with-proof-omitted} below thus provides us
with:
\begin{gather*}
  \dim_\K\! {\mathcal V}_d=(d+1)(D_X+1)((d+1)D_Y+1)-\binom{d\Delta+\Delta+1}2,\\
  \dim_\K\! {\mathcal W}_d=\bigl((D_X+1)d+1\bigr)(D_Yd+1)-\binom{d\Delta}2+1.
\end{gather*}
Injecting the three dimensions in Eq.~\eqref{ineq-dimb} finally gives
the following condition:
\begin{multline*}
  d^2-\left(2D_XD_Y+D_Y-\Delta^2-\Delta-1\right)d\\
  {}+\frac{(\Delta-1)\Delta}2-(D_X+1)(D_Y+1)>0.
\end{multline*}
Call~$\psi$ the value obtained on substituting the bound of the
theorem for~$d$ in the left-hand side:
$$ \psi = (3Dy-1)Dx+Dy-3\Delta(\Delta+1)/2 +2.$$
{}From $D_Y\ge 1$ and $0\leq\Delta\leq\min(D_X,D_Y)=:m\geq1$ it follows
that
$$ \psi \ge (3m-1)m+m-3m(m+1)/2+2 = 3m(m-1)/2 +2 > 0.$$
\begin{lemma}\label{lem:combinatoric-with-proof-omitted}
  The number of monomials in $X$ and~$Y$ of total degree at most
  $\delta$, degree in~$X$ at most $\delta_X$, and degree in~$Y$ at
  most $\delta_Y$ is
  $(\delta_X+1)(\delta_Y+1)-\binom{\delta_X+\delta_Y-\delta+1}2$, 
  whenever $\delta_X + \delta_Y +1\ge \delta\ge\max(\delta_X,\delta_Y)$.
\end{lemma}
\subsection{Construction of Differential Operators} \label{algorithms}
\label{efficient-algo}
Given a polynomial $P \in \K[X,Y]$ satisfying~(H), we have
proved in the previous sections the existence of several operators in
$\K[X] \langle \Dx \rangle$ associated to $P$.  The aim of this
section is to study the efficient construction of such operators. This
is done under an additional, but mild, hypothesis ((H') below). As
a by-product, we obtain in Thm.~\ref{th:lift+PH} an alternative unified
proof of slightly weaker versions of
Thms.~\ref{tm:cockle},~\ref{th:simple-bound}
and~\ref{thm:bidegree-bound}.
\paragraph*{Direct Algorithms}
The proof of Thm.~\ref{th:simple-bound} boils down to a kernel
computation of a matrix of size $\sim \! D_X D_Y^{3}$. Even using FFT
for (univariate and bivariate) polynomial multiplication and a
Wiedemann-like approach for the linear algebra step, this computation
requires $\sim \! D_X^2 D_Y^{6}$ operations in $\K$.
\par
In the sequel we provide an algorithm with better cost, which is based
on Pad\'e--Hermite approximation of series expansions of an algebraic
solution and its derivatives.
\paragraph*{Notation and Assumptions}
We assume in the rest of this section that the following hypothesis
holds.
\begin{hypothesis}[{H'}] Besides Hypothesis~(H), 
  \begin{enumerate}
  \item[$(H_a)$] $D_Y \ge 2$ and $P$ is absolutely irreducible
    in~$\K[X,Y]$ (which means that $P$ is irreducible when seen over
    the algebraic closure $\KK$ of $\K$), and
  \item[$(H_b)$] $x=0$ annihilates neither the discriminant $\Delta_Y(X)$ nor the leading coefficient
    $P_{D_Y}(X)$ of $P$, seen in
    $\K[X][Y]$.
  \end{enumerate}
\end{hypothesis}
To clarify the status of these assumptions, note that $(H_a)$ is a
global property of $P$ which is crucial in the subsequent proof that
our algorithm works correctly. In contrast, the local property $(H_b)$
is purely of technical nature, being similar with that on \emph{lucky
  points} in Section~\ref{algo-cockle}.  $(H_b)$ can be ensured
deterministically and in good complexity after a change of coordinates
$X \gets X+a$ with $a\in \K$ not a zero of $P_{D_Y} \Delta_Y$.
\paragraph*{Principle of the Algorithm} 
Under $(H_b)$, the polynomial $p(Y):=P(0,Y)$ is separable of
degree $D_Y$. By Hensel's lemma, the $D_Y$ roots $\alpha_i$
of $P(X,\alpha_i)=0$ are all in $\KK[[X]]$.  To avoid the
factorization of $p(Y)$, we  work in the quotient algebra
$\A:=\K[Y]/(p(Y))$.  Let $y$ denote the residue class of $Y$ in $\A$.
By Hensel's lemma again, there exists a unique series $\varphi \in
\A[[X]]$ such that $\varphi(0)=y$ and $P(X,\varphi(X))=0$. Moreover,
the expansion of $\varphi$ to any prescribed precision can be
efficiently computed by 
Newton iteration.
\par
To give a unified presentation, we denote by $\partial$ one of the
derivation operators $\Dx$ and $\Tx$.  The principle of our algorithm
is to compute enough terms of $\varphi(X)$, so that the coefficients
of an associated operator can be recovered by Pad\'e--Hermite (abbrev.
PH) approximation of the vector $(\varphi, \partial \varphi,
\partial^2\varphi,\ldots)^t$ of the first derivatives of~$\varphi$
wrt $\partial$.  Thus, we obtain first an operator with
coefficients in $\A[X]$, from which we extract an operator in
$\K[X]\langle \partial \rangle$ associated to~$P$. We give in
Figure~\ref{fig:algtodiff} a detailed presentation of the algorithm.
The main difficulty of this approach is to provide suitable bounds on
the orders of truncation. This is done in the next theorem, which
encapsulates the main results of this section.
\begin{theorem}\label{th:lift+PH}
  Let $P \in \K[X,Y]$ be of bi-degree
  $(D_X,D_Y)$ and satisfy Hypothesis $(H')$. 
The number of ops in~$\K$ for
  \begin{enumerate}
  \item {\sf AlgToDiff}$(P, 4 D_X D_Y^2, D_Y,\partial)$ is $\bigOsoft (D_X D_Y^{\omega +3})$;
  \item {\sf AlgToDiff}$(P,5D_X D_Y,5 D_Y,\partial)$ is $\bigOsoft (D_X D_Y^{\omega +2})$;
  \item {\sf AlgToDiff}$(P,B,B,\partial)$ with $B=4D_X D_Y + D_Y -2D_X
    -2$ is $\bigOsoft (D_X^{\omega +
    1} D_Y^{\omega +2})$.
  \end{enumerate}
All three output a
    non-zero differential operator in $\K[X] \langle \partial \rangle$
    associated to $P$.
\end{theorem}
The first algorithm returns an operator of order at most~$D_Y$, but with coefficients of cubic degree. The second one returns an operator whose order is still linear in~$D_Y$, but whose coefficients have only quadratic degree. Finally, the last one returns an operator of quadratic order and degree, whose degree is smaller than the previous one.
\begin{figure}[t]
  \begin{center} 
    \fbox{\begin{minipage}{7.9cm}
  \begin{center}\textsf{AlgToDiff}($P,B_X,B_\partial,\partial$) \end{center}
      \textbf{Input:} $P \in \K[X,Y]$, $B_X, B_\partial \in
          \mathbb{N}$, $\partial \in \{\Dx,\Tx \}$
          \par\smallskip \textbf{Output:} an operator  ${L} \in
          \K[X] \langle \partial
          \rangle\setminus\{0\}$~associated~to~$P$, with $\deg_X(L)
          \le B_X$ and $\deg_\theta(L) \le B_\partial$.
        \begin{tabbing}
           1. Set $\Sigma:=B_XB_\partial + B_X + B_\partial $. \\
           2. Expand  $\varphi(X) \in \A[[X]]$ up to  precision
           $\Sigma + B_\partial$. \\
           3. Compute   $Z_i = \partial^i \varphi$, for
           $i=0,1,\ldots,B_\partial$ to precision $\Sigma$.
  \\
            4. Compute a PH approximant $\big(\ell_0(X), \ldots,
  \ell_{B_\partial}(X) \big) \neq 0$ \\ \quad in $\A[X]^{B_\partial+1}$ 
               of $(Z_0,\ldots,Z_{B_\partial})$, of type
  $(B_X,\ldots,B_X)$.  \\ 
             5. Write   ${L} =  \ell_0(X) + \ell_1(X) \partial +
  \cdots + \ell_{B_\partial}(X) \partial^{B_\partial}$ as \\
                $\quad {L}_0 + y {L}_1 + \cdots + y^{D_Y -1}
  {L}_{D_Y -1},$ with ${L}_i \in \K[X]
                \langle \partial \rangle$.  \\ 
               6. Return any ${ L}_i \neq 0$.
        \end{tabbing}
      \end{minipage}
    }\end{center}
  \caption{Construction of differential operators by Pad\'e--Hermite
  approximation.}
  \label{fig:algtodiff}
\end{figure} 

First we give a bound on the number of terms needed to reconstruct an operator.
\begin{lemma} \label{lemma:one-truncated-solution-implies-associated}
  Under~(H'), let $L \in \KK[X]\langle \partial \rangle \setminus
  \{0\}$ be of order at most $B_\partial$ in $\partial$, with
  coefficients of degrees at most $B_X$.
  \begin{enumerate}
  \item[(a)] If $\alpha_0 \in \KK[[X]]$ is such that $P(X,\alpha_0)=0$
    and $L \alpha_0=0$, then $L$ is associated to $P$.
  \item[(b)] Let $\sigma = 4D_X D_Y B_\partial + B_X D_Y -2D_X
    B_\partial$. If $\alpha_0 \in \KK[[X]]$ is such that
    $P(X,\alpha_0)$ and $L \alpha_0$ are zero modulo $X^\sigma$, then
    $L$ is associated to $P$.
  \end{enumerate}
\end{lemma}
\begin{proof}
  There exists $Q(X,Y) \in \KK[X,Y]$ with
  \begin{align*}                 
    \deg_X(Q) &\le 2 B_\partial D_X +B_X - D_X\\
    \deg_Y(Q) &\le 2(D_Y-1) B_\partial - D_Y+2,
  \end{align*}
  such that $L \alpha = \frac{Q(X,\alpha)}{P_Y(X,\alpha)^{2 B_\partial
      - 1}}$ for any root $\alpha \in \overline{\K(X)}$ of
  $P(X,\alpha)=0$.  Indeed, if $L=\ell_0(X) + 
  \cdots + \ell_{B_\partial}(X) \partial^{B_\partial}$, we take $Q = $
  $\ell_0 Y P_Y^{2B_\partial-1} + \ell_1 W_1 P_Y^{2B_\partial-2}
  +\cdots + \ell_{B_\partial} W_{B_\partial}$, where the $W_i(X,Y)$
  are the polynomials introduced in Section~\ref{resolvent}.
  \par
  Let $R(X) \in \KK[X]$ be the resultant of $P(X,Y)$ and $Q(X,Y)$ with
  respect to $Y$.  To show that $L$ is associated to $P$, it is enough
  to prove that $R$ is identically zero. Indeed, since $P$ is
  irreducible in $\KK[X,Y]$, the nullity of $R$ implies that $P$
  divides $Q$~\cite[Cor.~6.20]{GaGe03}, therefore $Q(X,\alpha)=0$, and
  thus $L\alpha =0$, for any $\alpha \in \overline{\K(X)}$ such that
  $P(X,\alpha)=0$.
  \par
  $(a)$ By definition, there exist polynomials $U,V \in \KK[X,Y]$ such
  that $U(X,Y) P(X,Y) + V(X,Y) Q(X,Y) = R(X)$. Evaluating this
  equality at $Y=\alpha_0$ yields $R(X) = 0$, as wanted. The same
  reasoning shows that under the hypothesis of $(b)$ we have $R(X) = 0
  \bmod X^\sigma.$ On the other hand, by B\'ezout's theorem, one can
  bound the degree of the resultant $R$ by $D_X \deg_Y Q + D_Y \deg_X
  Q$. Therefore $\deg(R) $ is at most
  \begin{multline*}
    D_X (2D_Y B_\partial - 2B_\partial -D_Y + 2) + D_Y (2B_\partial D_X +
    B_X - D_X) \\ < 4D_X D_Y B_\partial + B_X D_Y -2D_X B_\partial =
    \sigma,
  \end{multline*}
  since $D_Y \ge 2$. Finally $\deg(R) \leq \sigma$ and $R(X) = 0 \bmod
  X^\sigma$ imply that $R$ is the zero polynomial.
\end{proof}
\begin{proof}[of Thm.~\ref{th:lift+PH}]
  In Step 4, the Pad\'e--Hermite equality $\sum_i \ell_i Z_i = 0 \bmod X^\Sigma$
   can be viewed as an under-determined homogeneous linear
  system with $\Sigma$ equations in the $\Sigma+1$ unknown
  coefficients of the polynomials $\ell_i$. It thus admits at least
  one non-trivial solution in $\A^{\Sigma + 1}$.
  \par
  Next let $p=\prod_i p_i$ represent the irreducible factorization of
  $p$ in $\K[Y]$, and let $\A_i= \K[Y]/(p_i(Y))$, so that $\A$
  rewrites into the product of fields $\prod_i \A_i$. Let
  $\varphi^{(i)}$ be the image of $\varphi$ in $\A_i[[X]] \subseteq
  \KK[[X]]$ and let $L^{(i)}$ be the image of $L$ in $\A_i[X]\langle
  \partial \rangle \subseteq \KK[X] \langle \partial \rangle$. By
  construction $P(X,\varphi^{(i)}) = L^{(i)} \varphi^{(i)} = 0 \bmod
  X^\Sigma$ holds in $\KK[[X]$. Since all choices of $(B_X,
  B_\partial)$ as specified in Thm~4 ensure $\Sigma \geq \sigma$, it
  follows from
  Lemma~\ref{lemma:one-truncated-solution-implies-associated}(b) that
  $L^{(i)}$ is associated to $P$ for all~$i$. In other words, for any
  root $a \in \KK$ of $p(a) = 0$, and any root $\alpha \in \KK[[X]]$
  of $P(X,\alpha)=0$ we have that $L_0 \alpha + a L_1 \alpha + \cdots
  + a^{D_Y-1} L_{D_Y-1} \alpha =0$. We thus deduce that $L_i$ is
  associated to $P$ for all $i$, which concludes the proof of correctness.
  \par
  \smallskip We are left with the detailed description and complexity
  study of each step of the algorithm. Operations $(+,\times)$ in $\A$
  have cost $\bigO(\M(D_Y))$ operations in~$\K$, while inversion (when
  possible) takes $\bigO(\M(D_Y) \log D_Y)$. For simplicity, in what
  follows we suppose that FFT is used for polynomial multiplication,
  thus $\M(d) \in \bigOsoft (d)$ for all $d$.
  \par
  For the lifting step 2, we use the Newton iteration.
  By~\cite[Algo.~2, Prop.~3]{ChLe05}, under the hypothesis (H'), the
  first~$\Sigma$ terms of $\varphi \in \A[[X]]$ cost $\bigO(\Sigma
  {D_Y}^{(\omega+1)/2} + \sqrt{D_Y} \M(\Sigma) \M(D_Y)) \in \bigOsoft
  (B_X B_\partial D_Y ^{(\omega+1)/2})$ ops in $\K$.
  \par
  Since differentiating has linear complexity, Step 3 requires
  $\bigO(B_\partial^2 B_X)$ ops  in $\A$, hence at
  most $\bigOsoft(B_\partial^2 B_X D_Y)$ ops in $\K$.
  \par
For Step 5, we use the approximation
  algorithm in~\cite{Storjohann06}, which computes
  a nonzero Pad{\'e}--Hermite approximant of type $(B_X, \dots,B_X)$ for $(Z_0, \dots, Z_{B_\partial})$
   within $\bigOsoft\big(B_\partial^{\omega}
  B_X\big)$ ops in $\A$, provided that $\A$ is a field, or more
  generally, if no division by a non-invertible element in $\A$ occurs.
  \par
  In the general case, since $\A$ is not a field, we use the following
  strategy. We begin by applying the approximation algorithm over
  $\A$. If no forbidden division occurs, this algorithm outputs the
  desired operator $L$ with coefficients in $\A[X]$. If we encounter a
  nonzero $q \in \A$ that is not invertible, then by a gcd computation we obtain a proper
  factor $\tilde{p}$ of $p$ of degree at most $D_Y/2$ such that $q$ is
  either invertible or zero modulo $\tilde{p}$. This takes $\bigOsoft(D_Y)$
  operations in $\K$~\cite[Th.~11.5]{GaGe03}. We replace $\A$ by
  $\K[Y]/(\tilde{p}(Y))$ and restart the approximation algorithm. This
  procedure is repeated at most $\log D_Y$ times, until all the
  divisions are doable. The total cost of this computation is
  $\bigOsoft \big(B_\partial^{\omega} B_X D_Y\big)$ operations in
  $\K$.
  \par
  Finally, Steps 1, 5 and 6 do not use any arithmetic operation.
  Adding together these costs, and under the assumption $D_Y \leq
  B_\partial^2$ (satisfied by the choices in
  Thm.~\ref{th:lift+PH}), we get that
  \textsf{AlgToDiff}($P,B_X,B_\partial,\partial$) uses $\bigOsoft
  \big(B_\partial^{\omega} B_X D_Y\big)$ ops in $\K$.
\end{proof}
\paragraph*{Probabilistic and Heuristic Versions}
We now give a probabilistic version of
Algorithm {\sf AlgToDiff}, called {\sf AlgToDiffP}. The idea is to compute a Pad\'e--Hermite
approximant over~$\K$ rather than over $\A$. The probabilistic
algorithm {\sf AlgToDiffP} is obtained by replacing Steps 3--6 in
{\sf AlgToDiff} by: \\
$3'$. Let $\varphi(X) = \varphi_0(X) + y \varphi_1(X) + \cdots +
y^{D_Y-1} \varphi_{D_Y-1}(X)$.\\ \quad Choose random
$a_0,a_1,\ldots,a_{D_Y-1} \in \K\setminus \{0\}$ and compute \\ \quad
\; $Z_0 = \sum_i a_i \varphi_i(X) \bmod X^{\Sigma}$, $Z_i = \partial^i
Z_0, i=1,\ldots,B_\partial$. \\
$4'$. Compute a PH approximant $\big(\ell_0(X), \ldots,
           \ell_{B_\partial}(X) \big)$ in $\K[X]^{B_\partial+1}$ of
           $(Z_0,Z_1,\ldots,Z_{B_\partial})$ of type $(B_X,\ldots,B_X)$.  \\
$5'$. Return ${L} = \ell_0(X) + \ell_1(X) \partial + \cdots +
\ell_{B_\partial}(X) \partial^{B_\partial}.$

{\sf AlgToDiffP}$(P,5D_X D_Y, 5D_Y,\partial)$ uses $\bigOsoft(D_X D_Y^{(\omega +5)/2})$ ops in~$\K$; with $B$ as in Thm.~\ref{th:lift+PH}, the cost of {\sf
  AlgToDiffP}$(P,B,B,\partial)$ is $\bigOsoft (D_X^2 D_Y^{(\omega +5)/2} + (D_X D_Y)^{\omega + 1})$.
The exponent of~$D_Y$ is better than for the deterministic algorithms.
\par
Another improvement comes from the remark that the algorithms derived
from {\sf AlgToDiff} and {\sf AlgToDiffP} do not return operators of
order and degree as small as predicted by Thms.~\ref{th:simple-bound}
and~\ref{thm:bidegree-bound}. The reason for this is technical: we
need the values~$B_X,B_\partial$ it uses in order to prove that the
operator constructed by Pad\'e--Hermite approximation
cancels~$\varphi$ and not merely its truncation. Replacing arguments
$B_X,B_\partial$ by the better bounds from Thms.~\ref{th:simple-bound}
and~\ref{thm:bidegree-bound} we obtain heuristic modifications {\sf
  AlgToDiffP}$(P,3D_X D_Y+6D_Y, 6D_Y,\partial)$ and {\sf
  AlgToDiffP}$(P,B,B,\partial)$ with $B$ as in
Thm.~\ref{thm:bidegree-bound}. So far, we do not have a proof that
these heuristics always return a correct output, but they behave very
well in practice (see \S\ref{sec:experiments}).
\subsection{Unrolling the Recurrence}  \label{recurrence}
The operators from Thm.~\ref{th:lift+PH} let
compute  series expansions of algebraic
functions asymptotically faster than Newton iteration. The following quantifies the improvement.

\begin{theorem}\label{theo:unroll} 
Let $\psi\in\LL[[X]]$ be such that $P(X,\psi(X))=0$, where $\LL=\K(\alpha)$ is a finite extension of~$\K\subset\C$. Then, under~(H'), the first $N$ terms of $\psi$
can be computed
in\\ $\bigO\left(N D_X D_Y([\LL:\K]+\M(D_Y)/D_Y)\right) $ ops in $\K$, as $N\rightarrow\infty$.
\end{theorem}
In the case when~$\LL=\K$, the complexity we get is therefore linear in~$N$ and \emph{quadratic} in the degree if fast multiplication is used.
A better complexity~$O(ND)$ was announced by Chudnovsky \& Chudnovsky in~\cite{ChCh86,ChCh87}. While the dependency in~$N$ is correct, they seem to have overlooked the degree growth of the coefficients of the differential resolvent, that gives a cubic dependency in~$D$ for their algorithm.

Our algorithm is presented in the proof below. 
\begin{proof}
The first part of the computation is independent of~$N$. 
  We begin by calling {\sf AlgToDiff}$(P,B_X,B_\partial,\partial)$
  with $B_X=5D_X D_Y$, $B_\partial =5D_Y$ and $\partial =\Tx$. We then
  convert the operator into a recurrence of the form $$r_0(n) u_{n} +
  r_1(n) u_{n+1} + \cdots + r_{B_X}(n) u_{n+B_X} = 0, \quad \text{for}
  \; n\geq 0 $$ with coefficients $r_i(X) \in \K[X]$ of degree at
  most~$B_\partial$.
Let~$\rho$ be the largest nonnegative integer root of the polynomial $r_{B_X}(X)$, or~$-1$ if no such root exists. Thus for any $n>\rho$, $u_{n+B_X}$ can be computed from the previous ones using the recurrence. 
 Note that we also have at our disposal the
  initial segment $\sum_{n=0}^{B_X-1} u_n X^n$ of $\varphi(X)$.
For practical purposes $\rho$ can be replaced by an upper bound like a Cauchy bound (see, e.g.~\cite[Lemma~6.7]{Yap00}).
If~$\rho\neq-1$, we use Newton iteration to extend this initial segment up to the index~$\rho+B_X$ if necessary. Using the minimal polynomial of~$\alpha$, we then deduce the corresponding segment of~$\psi$.

Only the rest of the computation depends on~$N$.
The recurrence is unrolled
for $n$ from $\rho+1$ to $N-B_X-1$. 
Since the coefficients~$r_i$ belong to~$\K$, so do the values~$r_i(n)$ and thus
each new coefficient $u_n$ of
  $\psi$ can be obtained using $\bigO(B_X[\LL:\K]) \in \bigO(D_X
  D_Y[\LL:\K])$ operations in $\K$, provided that the values taken by the
  coefficients $r_i(X)$ at the points $\rho+1,\ldots, N-B_X-1$ have been
  computed and stored. 

This computation can be done separately for
  each coefficient $r_i(X)$ using fast multipoint evaluation on
  $\rho+1,\rho+2,\dots,\rho+B_\partial$ and then fast extrapolation on arithmetic
  sequences. The cost of all the fast multipoint evaluation steps is
  $\bigO(B_X \M(B_\partial) \log B_\partial)$ and that of all the fast
  extrapolation steps is $\bigO(B_X N \M(B_\partial)/B_\partial)$, see
  e.g.~\cite{GaGe03,BoGaSc07}.
\end{proof}
We note that Hypothesis (H') is not strictly necessary in practice: the same method handles singular cases as well, provided power series solutions exist at the point of interest.
\section{Experiments} \label{sec:experiments}
\subsection{Timings}
We have implemented the algorithm of \S\ref{algo-cockle} in 
\texttt{Magma} 
V2.11-14~\cite{Magma97}. In
Table~\ref{tb:cockle} we report on our experiments with random dense
bivariate polynomials over the prime field with $9973$ elements on a
Pentium (M) 1.8GHz processor. Column ``ser.'' indicates the time
(in sec.) of our algorithm, while column ``rat.'' is for
\texttt{Magma}'s 
\texttt{DifferentialOperator}, where we
removed the characteristic $0$ requirement. The latter routine
implements Cockle's algorithm in $\K(X)$ and has been contributed by
A.~van der Waall. Column ``$\eta$'' indicates the value of $\eta$
and column ``$\deg_X(M)$'' gives the actual degree
in $X$ of the minimal operator. We observe that Cockle's algorithm is
always faster when run over the series and that there is still room
for improvement on our degree bound $\eta$.
\par
In Table~\ref{tb:algtodiff}, we compare the practical behaviour of our
heuristic algorithm from~\S\ref{efficient-algo} for 
algebraic
series defined by polynomials of bi-degree $(1,D_Y)$ with that of
Newton's iteration (column \textbf{Newton}). Column \textbf{AlgToRec}
contains the cumulated timings of the heuristic \textsf{AlgToDiffP}$(P,B,B,\partial)$ (with $B$ as in Th.~\ref{thm:bidegree-bound}) and of
the conversion of a differential operator into a recurrence. Column
$N$ gives the number of terms of the expansion, while column
\textbf{unroll} shows the timings for unrolling the recurrence.  One
motivation for the computation when~$D_X=1$ is its
application in univariate polynomial root finding by
homotopy in~\cite[\S~10]{ChCh90}.

These experiments show that there is an initial cost due to the size of the recurrence being constructed. Once the number of desired coefficients is high enough compared to this size, then the method is very effective.
\begin{table}[t]
  \caption{Cockle's algorithm}
  \label{tb:cockle}
  \begin{center}
    \begin{tabular}{ccccc}
      \hline
      $(D_X,D_Y)$ & ser. & rat. & $\eta$ & $\deg_X(M)$\\ 
      \hline
      (1,1) & 0.002 & 0.002 & 2 & 2\\ 
      (2,2) & 0.003 & 0.004 & 17 & 10\\ 
      (3,3) & 0.02 & 0.03 & 69 & 36\\ 
      (4,4) & 0.10 & 0.16 & 182 & 92\\ 
      (5,5) & 0.47 & 0.98 & 380 & 190\\ 
      (6,6) & 1.86 & 4.56 & 687 & 342\\ 
      (7,7) & 5.80 & 16.5 & 1127 & 560\\ 
      (8,8) & 15.5 & 49.9 & 1724 & 856\\ 
      (9,9) & 38.0 & 138 & 2502 & 1242\\ 
      (10,10) & 72.7 & 340 & 3485 & 1730\\ 
      \hline
    \end{tabular}
  \end{center}
\end{table}
\begin{table}[t]
  \caption{{Newton's} algorithm vs. {AlgToDiffP} }
  \label{tb:algtodiff}
  \begin{center}
    \begin{tabular}{ccccc}
      \hline
      $(D_X,D_Y)$ & $N$ & \text{AlgToRec} & \text{unroll} & \text{Newton} \\
      \hline
      (1,8) & $2^{10}$ & 6.66 & 0.22 & 1.75 \\
      (1,8) & $2^{11}$ & 6.66 & 0.44 & 3.91 \\
      (1,8) & $2^{12}$ & 6.66 & 0.89 & 8.83 \\
      (1,8) & $2^{13}$ & 6.66 & 1.8  & 18.84 \\
      (1,8) & $2^{14}$ & 6.66 & 3.67 & 42.53 \\
      (1,8) & $2^{15}$ & 6.66 & 7.38 & 94.99 \\   \hline
      (1,9) & $2^{10}$ & 10.54 & 0.26 & 2.14 \\
      (1,9) & $2^{11}$ & 10.54 & 0.44 & 4.35 \\
      (1,9) & $2^{12}$ & 10.54 & 1.07 & 9.31 \\
      (1,9) & $2^{13}$ & 10.54 & 2.17 & 20.64 \\
      (1,9) & $2^{14}$ & 10.54 & 4.36 & 43.44 \\
      (1,9) & $2^{15}$ & 10.54 & 8.75 & 97.64 \\

      \hline
    \end{tabular}
  \end{center}
\end{table}
\subsection{On the Optimality of the Bounds}  \label{optimal-bounds}
\paragraph*{Conjectures}       
Intensive experiments suggest that the degree of the differential resolvent is
bounded by $D(D^2-5D/2+5/2)$ for polynomials of total degree $D$ and
$D(2D^2-3D+3)$ for polynomials of bi-degree $(D,D)$, within a factor~2 of
the bound of Thm.~\ref{tm:cockle}.
Also,  the bound from
Thm.~\ref{thm:bidegree-bound} seems only slightly pessimistic: for a
polynomial of total degree $D$, there often exists a recurrence of
order $D^2-2 $ with coefficients of degree at most $D^2-1$.  For a
polynomial of bi-degree $(D_X,D_Y)$, experiments suggest the bound
$2D_X D_Y-2 - (D_X-D_Y)$ if $D_Y>1$ and $D_X + 1$ if $D_Y=1$.
\paragraph*{Lower Bound}                        
We note also that $D (D-1)$ is the degree of the discriminant of a generic polynomial~$P$ of total degree~$D$.
Generically, this discriminant is square-free. The differential equation
satisfied by all solutions of~$P$ has to be singular at the roots of the discriminant
and therefore its leading coefficient cannot have a smaller degree.
Thus our bound on the order of the recurrence is asymptotically
optimal.

An example illustrating $D(D-1)$ as a lower bound is provided by
$P(X,Y) = Y^D - Y + X^D$. Then $\alpha = X^D + X^{D^2} +
\bigO(X^{D^2})$ is a root of $P$ seen in $\Q[[X]][Y]$.  It has a
sequence of $D(D-1)-1$ zero coefficients before a non-zero one. Thus,
the order of the recurrence satisfied by the coefficients of $\alpha$
is at least $D(D-1)$.
\section*{Acknowledgments}
This work was supported in part by the French Research Agency (ANR Gecko).
%

%
\end{document}